\documentclass[twocolumn,floatfix,pra,aps,showpacs,superscriptaddress]{revtex4}
\usepackage{epsfig,graphicx}
\usepackage{flafter}
\usepackage{amsmath}
\usepackage{color}
\usepackage{bm}
\usepackage{amssymb}
\usepackage{epstopdf}
\usepackage{amstext}
\usepackage{amsthm}
\usepackage{amsfonts}
\usepackage{latexsym}
\usepackage{multirow}
\usepackage{mathtools}
\usepackage{bm}
\usepackage{bbm}
    \usepackage{amsmath}
    \usepackage{amsthm}
    \usepackage{amsfonts}
    \usepackage{braket}
\usepackage{soul}

\begin{document}

\title{Rotating a supersolid dipolar gas} 

\author{S. M. Roccuzzo}
\affiliation{INO-CNR BEC Center and Dipartimento di Fisica, Universit\`a degli Studi di Trento, 38123 Povo, Italy}
\affiliation{Trento  Institute  for  Fundamental  Physics  and  Applications,  INFN,  38123,  Trento,  Italy}
\author{A. Gallem\'{\i}}
\affiliation{INO-CNR BEC Center and Dipartimento di Fisica, Universit\`a degli Studi di Trento, 38123 Povo, Italy}
\affiliation{Trento  Institute  for  Fundamental  Physics  and  Applications,  INFN,  38123,  Trento,  Italy}
\author{A. Recati\footnote{Corresponding Author: alessio.recati@ino.it}
}
\affiliation{INO-CNR BEC Center and Dipartimento di Fisica, Universit\`a degli Studi di Trento, 38123 Povo, Italy}
\affiliation{Trento  Institute  for  Fundamental  Physics  and  Applications,  INFN,  38123,  Trento,  Italy}
\author{S. Stringari}
\affiliation{INO-CNR BEC Center and Dipartimento di Fisica, Universit\`a degli Studi di Trento, 38123 Povo, Italy}
\affiliation{Trento  Institute  for  Fundamental  Physics  and  Applications,  INFN,  38123,  Trento,  Italy}

\date{\today}

\begin{abstract}
Distinctive features of supersolids show up in their rotational properties. 
We calculate  the moment of inertia of a harmonically trapped dipolar Bose-Einstein condensed gas 
as a function of the tunable  scattering length parameter, providing the transition from the (fully) superfluid to the supersolid phase and eventually to an incoherent crystal of self-bound droplets. The transition from the superfluid to the supersolid phase is characterized by a jump in the moment of inertia, revealing its first order nature. In the case of elongated trapping in the plane of rotation we show that the moment of inertia determines the  value of the frequency of the scissors mode, which is  significantly affected by the reduction of superfluidity  in  the supersolid phase.
The case of an in-plane isotropic  trapping is instead well suited to study the formation of quantized vortices, which are shown to be characterized, in the supersolid phase, by a sizeable deformed core, caused by the presence of the surrounding density peaks.
\end{abstract}

\maketitle

The study of the rotational behavior of a many-body system provides a crucial  test to identify 
the effects of superfluidity. This test became particularly important when experimentalists tried 
to investigate the possible signature of superfluidity in a crystal of solid $^4$He \cite{Chan1}, 
looking for deviations of the moment of inertia from the classical rigid body value by means of a 
torsional oscillator. These experiments were later shown  to be inconclusive in providing evidence 
for the long sought effect of supersolidity \cite{Balibar,Chan2}. Ultracold atoms have eventually 
proved to be more efficient platforms. In 2017 two experiments reported 
on the first creation of supersolidity employing Bose-Einstein condensates inside optical 
resonators \cite{ETH_SS1,ETH_SS2} and spin-orbit coupled mixtures \cite{MIT_SS}. The teams of 
Florence \cite{F1}, Stuttgart \cite{S1} and Innsbruck \cite{I1} have later observed supersolid 
properties in a harmonically trapped dipolar Bose-Einstein condensate, revealing at the same time 
the effects of the crystal modulation of the density profiles and the ones  of coherence. 
More recent works of the same teams \cite{F2,S2,I2} measured the Goldstone modes 
associated with the spontaneous breaking of the relevant symmetries characterizing the supersolid 
phase. On the theoretical side much work has been devoted in the past to the description of the 
supersolid phase of many-body systems and its superfluid behavior 
\cite{AndreevLifshitz,Saalow,Rica1994,Son,Dorsey2010}, emphasizing the peculiar features exploited 
by systems interacting with soft-core finite-range potentials 
\cite{Hydro-Rica2007,Boninsegni_SS2012,Ancilotto2013,Macri_SS}, the role of spin-orbit coupling 
\cite{Sandro_SO} and of long-range dipolar interactions  \cite{Roccuzzo1,Buechler2019,Pohl2019}. 

The purpose of the present Letter is to provide first  theoretical predictions  concerning  the 
rotational properties of a harmonically trapped supersolid dipolar gas, yielding unique information 
on the superfluid behavior of such a system, through the deviation  of its moment of inertia from 
the rigid value and the emergence of quantized vortices. Special emphasis will be given to the role 
played by the trapping potential which favors the direct observability of these relevant superfluid 
effects. The moment of inertia characterizes the global superfluid behavior of the non uniform 
system and can be easily calculated also in the presence of harmonic trapping and inhomogeneous 
configurations. 
In the presence of elongated trapping in the plane of rotation the moment of inertia dictates the value of the experimentally 
measurable frequency of the scissors mode, corresponding to an oscillating rotation of the gas. Isotropic trapping is instead well suited to host quantized vortices. These are predicted 
to exhibit a peculiar deformed core of large size, due to the strong reduction of the 
density in the interstitial region surrounding the high density peaks, which characterize the 
supersolid phase.
Interestingly, in the incoherent droplets state we always find a finite non-classical rotational inertia, 
due to the single droplet superfluidity.  

%

{\bf Moment of Inertia and the Scissors Mode}. 
The scissors mode~\cite{LoIudice,Lipparini} was first observed in nuclear physics \cite{RichterScissor}, 
where it consists of the relative 
oscillating rotation between neutrons and protons in deformed atomic nuclei. In atomic physics it 
was predicted \cite{Guery-Odelin} and soon measured \cite{Foot} in atomic Bose-Einstein condensates 
confined by an anisotropic external potential, confirming the typical irrotational behavior predicted 
by superfluidity. It was later studied and observed also in ultra-cold  Fermi gases~
\cite{TosiScissor,RudiScissor} and in 2D Bose gases \cite{PerrinScissor} as well as, more recently, 
in droplets of dipolar gases \cite{PfauScissor}. An easy estimate of the frequency 
of the scissors mode is provided by the sum rule approach~\cite{BecBook2016} based on the ratio
$ (\hbar \omega_{\rm sc})^2=m_1(L_z)/m_{-1}(L_z)$ between the energy weighted and the inverse energy weighted moments 
\mbox{$m_p(L_z)=\int d \omega \,\omega^p S(L_z,\omega)$} of the dynamic structure factor 
$S(L_z,\omega)= \sum_n |\bra{n}\hat{L}_z\ket{0}|^2\delta(\hbar \omega - \hbar \omega_n)$ relative 
to the angular momentum operator $\hat{L}_z$. 
Assuming the gas is confined in a harmonic potential 
$V_{\rm ho}({\bf r})=m/2 (\omega_x^2x^2+\omega_y^2y^2+\omega_z^2z^2)$,
the energy weighted moment takes the form (f-sum rule)
$2 m_1(L_z)=\langle[\hat{L}_z,[\hat{H},\hat{L}_z]\rangle = N{\hbar^2m}(\omega_y^2-\omega_x^2)\langle x^2-y^2\rangle$,
with $N$ and $m$ the atom number and the atom mass, respectively. 
It holds in general for central potentials commuting  with the angular momentum 
operator, and hence applies also to the case of  the anisotropic dipolar interaction, provided one chooses 
the component of the angular momentum  along the direction ($z$) of the dipole moments. 
In this case  the commutator
$[\sum_{ij}V_{dd}({\bf r}_{i}-{\bf r}_{j}),\hat{L}_z]$ vanishes identically, $V_{dd}({\bf r}_{i}-{\bf r}_{j})=\frac{\mu_0\mu^2}{4\pi}\frac{1-3\cos^2\theta}{|{\bf r}_{i}-{\bf r}_{j}|^3}$ being the dipole-dipole interaction between two identical magnetic dipoles aligned along the $z$-axis, 
$\theta $ the angle between the vector ${\bf r}_{i}-{\bf r}_{j}$ and the polarization direction $z$,  while $\mu$
is the atomic dipole moment and $\mu_0$ the vacuum permeability. 
The inverse energy weighted moment is instead related to the  moment of inertia per particle  $\Theta$
through the relation $2m_{-1}(L_z)=N\Theta$. 
It can be calculated  by applying the  static perturbation $-\Omega \hat{L}_z$ to the system 
and using the standard definition  $N\Theta=\lim_{\Omega \to 0}\langle\hat{L}_z\rangle/\Omega$.
Thus the frequency of the scissors mode takes the useful expression
\begin{equation}
(\hbar \omega_{\rm sc})^2 =  \frac {\hbar^2m(\omega_y^2-\omega_x^2)\langle x^2-y^2\rangle}{\Theta}.
\label{scissorsSR}
\end{equation}
\begin{figure}[t]
\includegraphics[width=\linewidth]{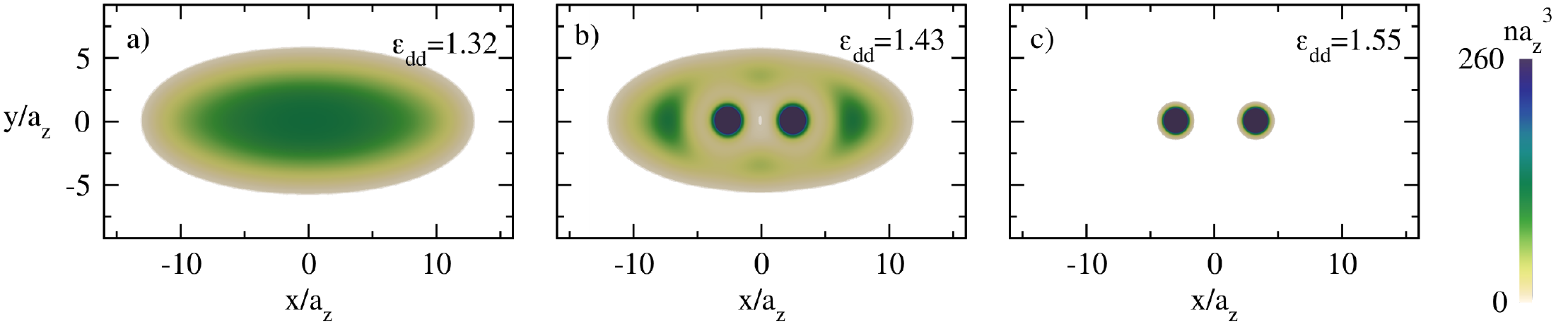}
\caption{Typical in-situ density profiles obtained from the stationary solution of the eGPE 
(\ref{gpequation}), for different values of $\varepsilon_{dd}$ (lengths are given in terms of 
the harmonic oscillator length $a_z=\sqrt{\hbar/m\omega_z}=0.87\mu m$). Three different regimes 
are clearly distinguishable: a) superfluid phase (for $\varepsilon_{dd}=1.32$), b) supersolid 
phase (for $\varepsilon_{dd}=1.43$), and c) an incoherent {\it crystal} of two self-bound droplets (for 
$\varepsilon_{dd}=1.55$). Notice that the color scale is saturated in (b) and (c), where the 
maximum value of the density reaches $na_z^3=900$ and $1800$, respectively.}
\label{fig:profiles1d}
\end{figure}

At zero temperature, 
the dipolar gas is characterized by a macroscopic wave function $\Psi({\bf r},t)$
that obeys the extended Gross-Pitaevskii equation (eGPE)~\cite{Wachter2016}
\begin{align}
&i\frac{\partial}{\partial t}  \Psi({\bf r},t)= \left[-\frac{\hbar^2}{2m}\nabla^2+V_{\rm ho}({\bf r})+g|\Psi({\bf r},t)|^2\right.\nonumber\\
+& \int d{\bf r'}V_{dd}({\bf r}-{\bf r'})|\Psi({\bf r'},t)|^2+\gamma(\varepsilon_{dd})|\Psi({\bf r},t)|^3\bigg]\Psi({\bf r},t),
\label{gpequation}
\end{align}
where the coupling constant $g=4\pi\hbar^2a/m$ is fixed by the  $s$-wave scattering length $a$ and 
$\varepsilon_{dd}=\mu_0\mu^2/(3g)=a_{dd}/a$ ($a_{dd}$ is the so-called dipolar length) is the ratio between the strength of the dipole-dipole 
and the contact interaction. The  last term is the local density approximation of the beyond-mean-field Lee-Huang-Yang (LHY) 
correction~\cite{FischerLHY,Pelster}, with $\gamma(\varepsilon_{dd})=\frac{16}{3\sqrt{\pi}} ga^{\frac{3}{2}}\int_0^{\pi}d\theta\sin\theta [1+\varepsilon_{dd}(3\cos^2\theta-1)]^{\frac{5}{2}}$.
The LHY term is  crucial in 
order to describe the supersolid phase and the occurrence of self-bound droplets. The use of the
LHY term in Eq.~(\ref{gpequation}) has been shown to work pretty well when 
compared with 
more microscopic (Monte Carlo) calculations \cite{Saito2016} and to properly capture the physics 
of the system when compared with experiments. The same has been shown to be the case 
for self-bound droplets of  quantum mixtures~\cite{PetrovDrop,TarruellDrop,TarruellDrop2,FattoriDrop,FattoriColl}.

First of all we determine the ground state configurations by evolving the eGPE in imaginary time
starting from a guess wave function. For the 
sake of concreteness we consider $N=4\times10^4$  $^{164}$Dy atoms confined in a harmonic potential 
with trapping 
frequencies equal to $\omega_{x,y,z}=2\pi\times(20,40,80)$ Hz. Such isotope has a dipolar length
$a_{dd}=131a_0$ ($a_0$ the Bohr radius) and it has been recently
shown to have a much longer lifetime with respect to the other magnetic atoms 
($^{162}$ Dy and $^{166}$ Er) ~\cite{I1}.

The eGPE admits solutions of different nature depending on $\varepsilon_{dd}$, 
which can be experimentally tuned by modifying the $s$-wave scattering length through Feshbach resonances.
We find that for $\varepsilon_{dd}<1.42$ the solution 
corresponds to a fully superfluid Bose-Einstein condensate (Fig. \ref{fig:profiles1d}(a)), with the shape of the density 
profile given by an inverted parabola \cite{Odell}. As  
$\varepsilon_{dd}$ increases the role of the dipolar interaction becomes more and more important and 
is at the origin of a rotonic structure in the excitation spectrum~\cite{SantosRoton,DellRoton}, 
observed experimentally in~\cite{IBKrotons}. The softening of the roton gap eventually causes 
the  transition to a density modulated structure: in the interval $1.42 < \varepsilon_{dd} <1.52$ 
the density profile of the equilibrium configuration  is characterized by typical overlapping density 
peaks, corresponding to the supersolid phase (see Fig. \ref{fig:profiles1d}(b)). For larger values of 
$\varepsilon_{dd}$ the density peaks do not overlap any more and form  an incoherent crystal of self 
bound droplets (see Fig. \ref{fig:profiles1d}(c)). 

\begin{figure}[t]
\includegraphics[width=\linewidth]{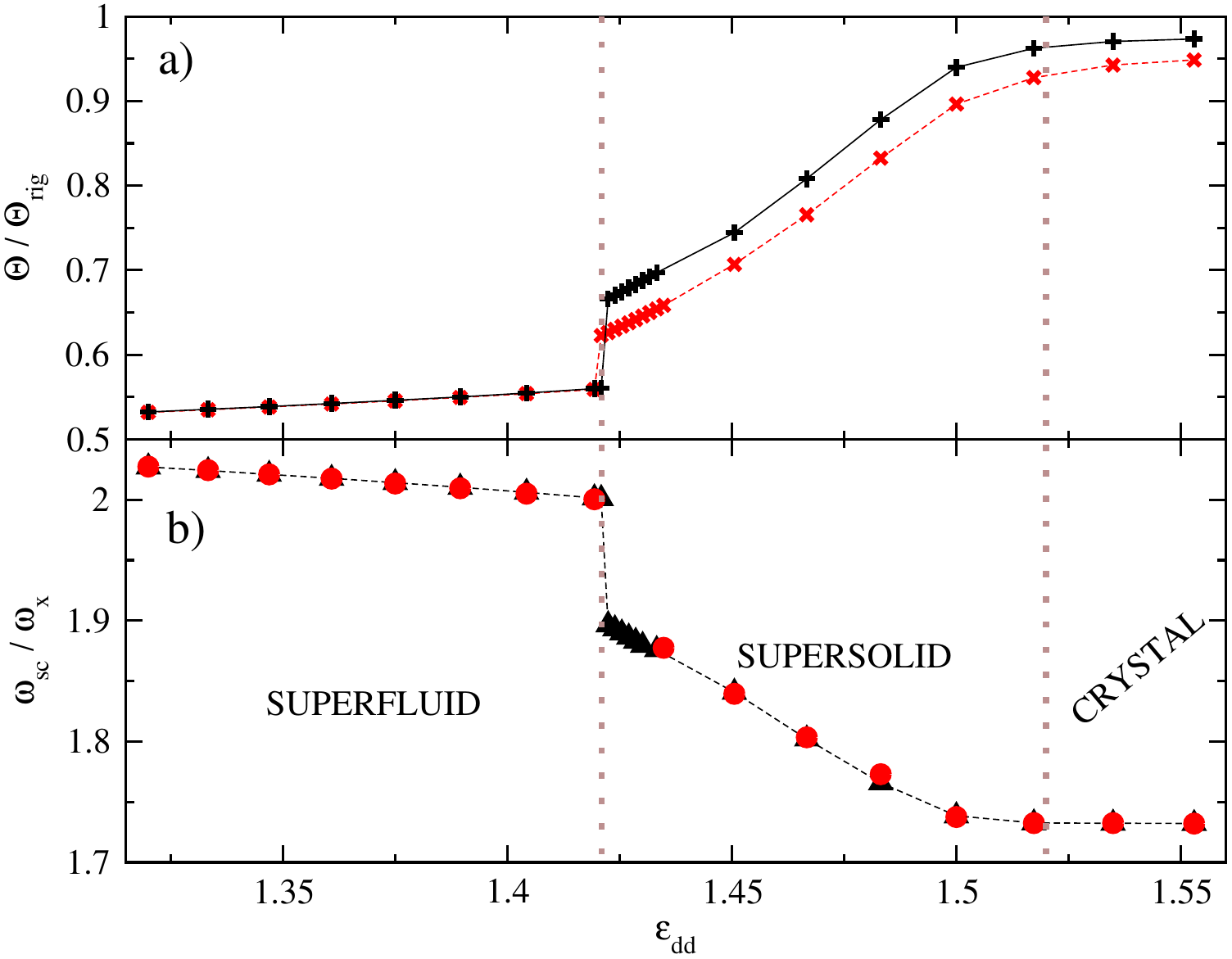}
\caption{a) Moment of inertia $\Theta$ of a dipolar gas in an axially-deformed harmonic trap 
($\omega_y/\omega_x=2$), as function of $\varepsilon_{dd}$. The black solid line shows the results 
of the eGPE calculations carried out with an angular momentum constraint and the red dotted line 
shows the estimate given by Eq. (\ref{ThetaBEC}). b) Frequency of the scissor mode, as function 
of $\varepsilon_{dd}$. The black dotted line corresponds to the sum rule estimate (\ref{scissorsSR}). 
Red circles correspond to the frequency of the time-dependent signal $\langle xy\rangle$ obtained 
from GP real-time simulations.}
\label{fig:I-1d}
\end{figure}

Once the phase diagram is known, we determine the moment of inertia by adding the term
$-\Omega \hat{L}_z\Psi$ to Eq. (\ref{gpequation}) and evaluating the angular momentum. Since the 
velocity field obtained within the eGPE has the irrotational form 
${\bf v}({\bf r})=\frac{\hbar}{m}\nabla S({\bf r})$, fixed by the gradient of the phase $S({\bf r})$ 
of the macroscopic wave function $\Psi({\bf r})=\sqrt{n({\bf r})}\exp(iS({\bf r}))$, this theory 
cannot describe a rigid rotational flow of the form ${\bf v}={\bf \Omega} \wedge {\bf r}$. 
Nevertheless, if the density profile is not rotationally invariant the moment of inertia can 
become large and even approach the rigid value, $\Theta_{\rm rig}=\int d{\bf r} (x^2+y^2)\,n({\bf r})$, 
as a consequence of the mechanical drag caused by the rotation. This is the case even in the fully 
superfluid phase if the trapping potential is highly elongated. It is, in fact, immediate to see 
that the variational result \cite{Fetter1974,BecBook2016}
\begin{equation}
\Theta_{\rm var} = \left(\frac{\langle x^2- y^2\rangle}{\langle x^2+  y^2\rangle}\right)^2\Theta_{\rm rig}
\label{ThetaBEC}
\end{equation}
for the moment of inertia, derivable 
making the ansatz $S=\alpha xy$ for the phase of the macroscopic wave function and 
satisfying the inequality $\Theta_{\rm var}\le \Theta$, approaches the 
rigid value for highly deformed configurations corresponding 
to $|\langle x^2- y^2\rangle|\simeq\langle x^2+  y^2\rangle$. 
In this extreme limit it is not possible to reveal the effects of superfluidity 
through the measurement of the scissors frequency Eq.~(\ref{scissorsSR}) and it is therefore convenient   
to work with moderately deformed traps. For this reason 
we have chosen the value $\omega_y/\omega_x=2$   
for the in-plane aspect ratio. It is worth noticing that in the fully superfluid phase the variational result  Eq.~(\ref{ThetaBEC}) coincides with the prediction of the hydrodynamic equations of superfluids \cite{Fetter1974,BecBook2016}. It can be also derived miroscopically in the case of the ideal Bose gas \cite{SandroInertia}. 
\begin{figure}[t]
\includegraphics[width=\linewidth]{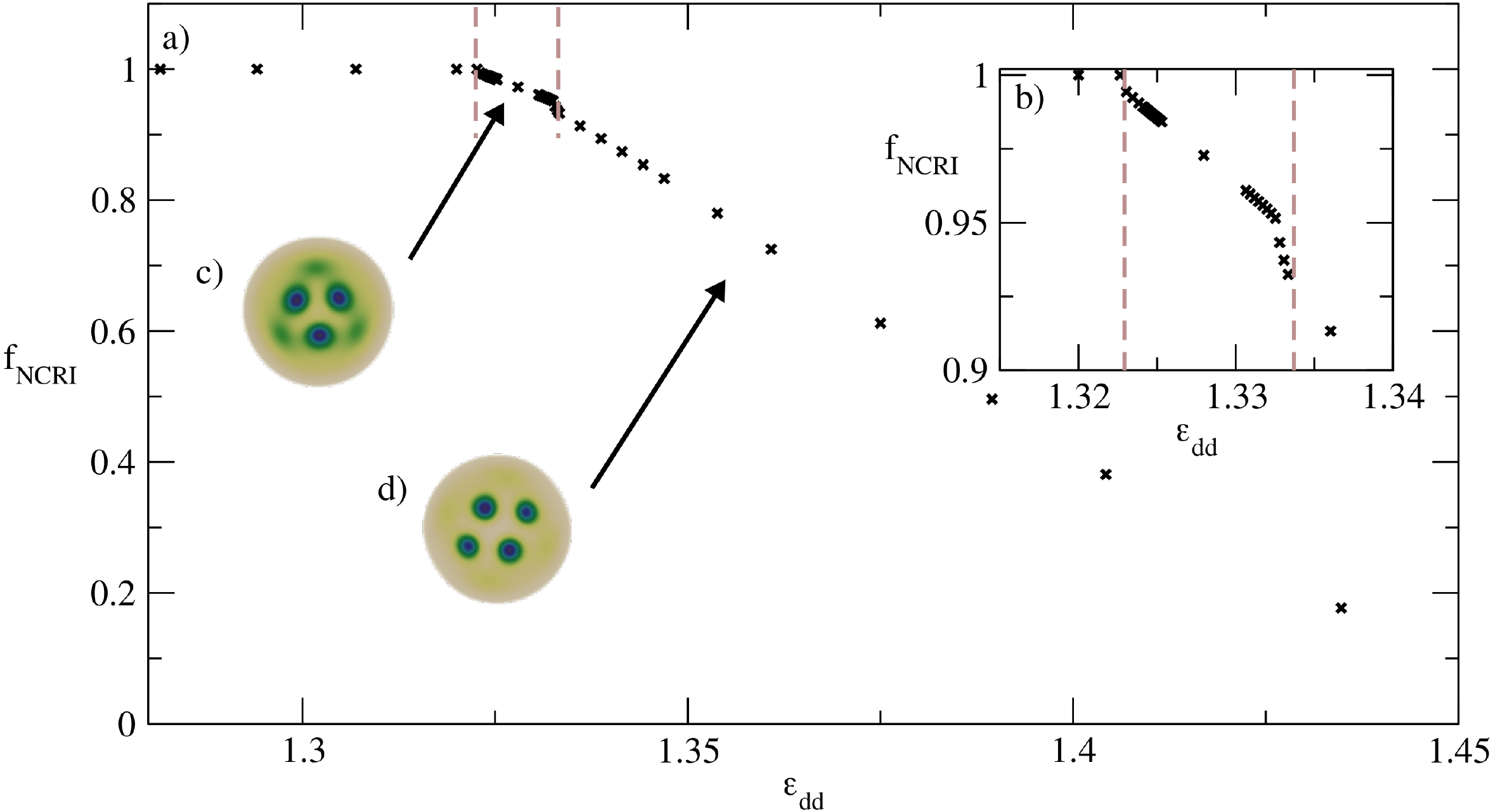}
\caption{a) Non classical rotational inertia fraction (\ref{eq:fs}) of a dipolar gas in an isotropic harmonic 
potential $\omega_y=\omega_x$ as a function of 
$\varepsilon_{dd}$. The brown dashed lines indicate the position of two jumps. b) Zoom in the 
region $1.315<\varepsilon_{dd}<1.34$, where the moment of inertia presents the two jumps. c) and d) Plot 
of the density in the region where the system presents a single-triangular cell and a two-triangular 
cell configuration, respectively.}
\label{fig:I-2d} 
\end{figure}

The results of our calculations for the moment of inertia
are reported in 
Fig. \ref{fig:I-1d}(a) in units of the rigid value. In the supersolid phase the ratio 
$\Theta/\Theta_{\rm rig}$ significantly increases  as a consequence of the presence of the density 
peaks which provides a solid-like contribution to $\Theta$. The transition between the superfluid 
and the supersolid phase is characterized by a visible jump that reflects its first order nature\cite{Rica1994,Boninsegni_SS2012,Macri_SS,Roccuzzo1,Lu2015}. 
By further increasing the value of $\varepsilon_{dd}$ the moment of inertia eventually approaches the 
rigid value, reflecting the crystalline nature of  the self-bound droplet phase. However, even in the crystal 
phase  the rigid body value is not exactly achieved since each droplet is itself 
superfluid and cannot host a rigid rotational motion. Because of  the small size of each 
droplet (compared to the inter droplet distance) as well as of  the anistropy of the trap, the 
difference between the rigid body value  and the one of the crystal phase is nevertheless almost 
negligible. In the elongated case reported in Fig.~\ref{fig:I-1d} it amounts to a few percent 
for the largest values of $\varepsilon_{dd}$. In Fig.~\ref{fig:I-1d} we also report (see red dotted 
line) the prediction of the approximate variational estimate (\ref{ThetaBEC}), which perfectly 
matches the numerical result in the fully superfluid  regime ($\varepsilon_{dd} <1.43$), while 
for larger values of $\varepsilon_{dd}$ it underestimates the actual value of the moment of inertia.

The moment of inertia can be used to estimate the frequency of the scissors mode, employing 
Eq.~(\ref{scissorsSR}). The predicted value ranges from the usual Bose-Einstein condensate value 
$\sqrt{\omega_y^2+\omega_x^2}=2.23\,\omega_x$ \cite{Guery-Odelin} for $\varepsilon_{dd} \to 0$,
to the value $\sqrt{\omega_y^2-\omega_x^2}=1.73\,\omega_x$ in the opposite 
limit of  large $\varepsilon_{dd}$, when $\langle y^2 \rangle\ll\langle x^2 \rangle$ and the 
moment of inertia takes the rigid value. The results for the frequency obtained within the sum 
rule approach  are reported as black triangles in Fig.~\ref{fig:I-1d}(b) as a function of 
$\varepsilon_{dd}$. In order to certify the validity of the sum rule 
prediction we have carried out real-time simulations of the eGPE by generating initially a 
sudden rotation of the confining trap. The relevant signal associated with the rotation of the 
cloud is provided by the quantity $\langle xy\rangle$. The simulation reveals the occurrence 
of a single well defined frequency for all values of $\varepsilon_{dd}$, which is in perfect 
agreement with the sum rule approach as shown by the red dots in Fig. \ref{fig:I-1d}(b). Measuring 
the jump in the frequency of the scissors mode at the superfluid-supersolid 
transition would provide an important proof of its first order nature. 

\begin{figure}
\includegraphics[width=\linewidth]{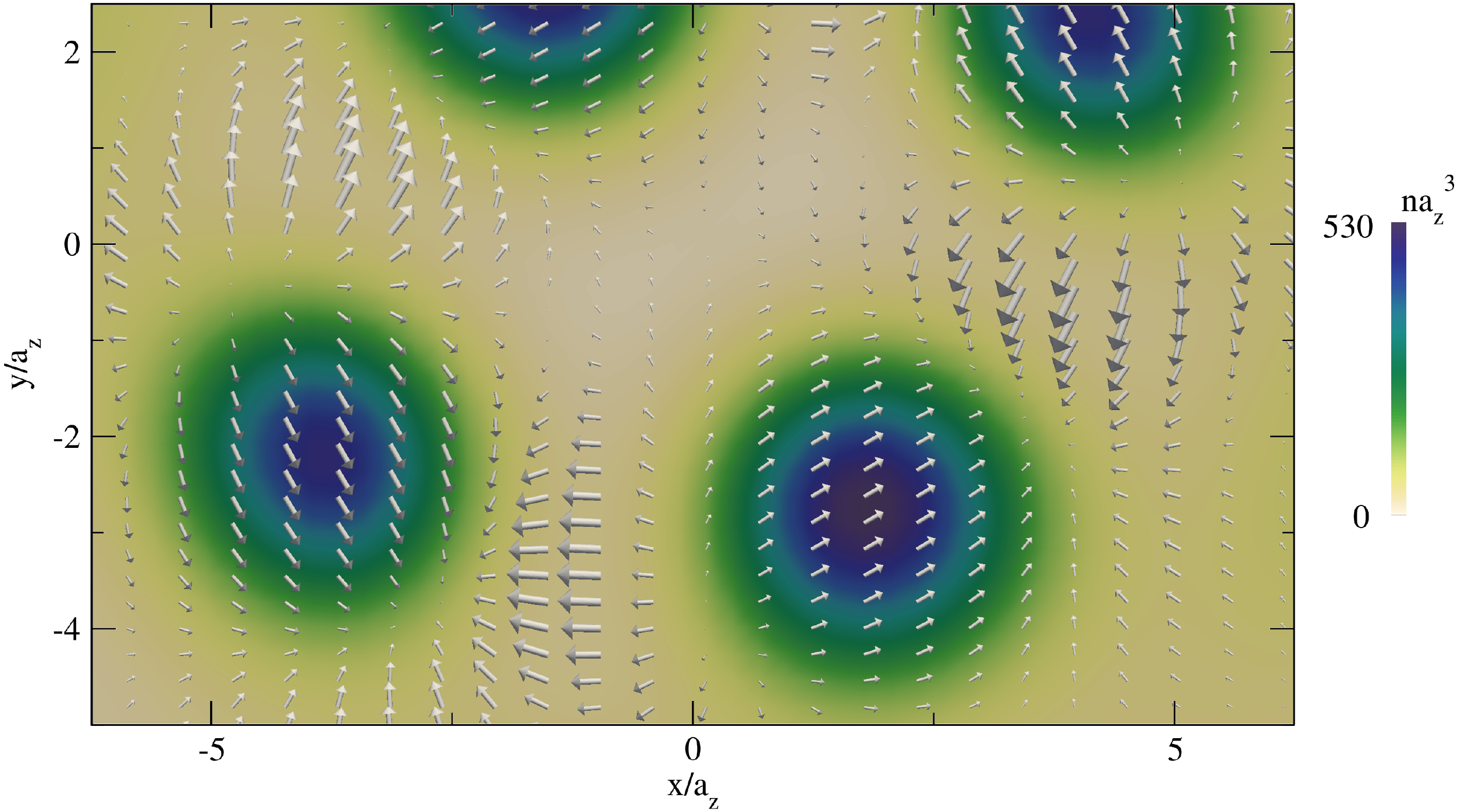}
\caption{Velocity field of a supersolid at small $\Omega$. The droplets partially follow the rigid body rotation, 
being dragged by the $-\Omega\hat{L}_z$ term of the hamiltonian. }
\label{fig:velocity}
\end{figure}

{\bf Non Classical Moment of Inertia and Quantized Vortices  in an Isotropic Trap}. If the 
confining potential is isotropic in the rotational plane ($\omega_x=\omega_y$) we find that, for 
$N=10^5$ $^{164}$Dy atoms and trapping frequencies equal to $\omega_{x,y,z}=2\pi\times(40,40,80)$ Hz, 
the formation of the supersolid phase emerges at the value $\varepsilon_{dd} =1.32$, with a density 
profile characterized by overlapping droplets arranged in triangular cells (see 
also \cite{Pohl2019}). The number and the distribution of the peaks depend on the atom number, the 
trapping frequencies and the scattering length. Nevertheless, the distance between droplets is 
essentially the same for all the configurations considered 
in the present work, and agrees rather well with the value of $2\pi/q_R=4.5a_z$ predicted 
in 2D uniform matter \cite{SantosRoton}, where the roton wave vector $q_R$ is determined
by the axial confinement length $a_z=\sqrt{\hbar/(m\omega_z)}=0.87 \mu m$.

In isotropic configurations the moment of inertia fixes the non classical rotational inertia (NCRI) 
fraction $f_{NCRI}$ according to the relation \footnote{The relationship between the moment of 
inertia and the NCRI fraction can be identified also  in the case of anisotropic confinement in the 
rotational plane by using the phenomenological relation 
$\Theta = (1-f_{NCRI})\Theta_{rig}+f_{NCRI} \beta^2\Theta_{rig}$, recently employed in \cite{TanziScissor}, 
where $\beta= \langle x^2-y^2\rangle /\langle x^2+y^2\rangle$ is  the deformation of the atomic cloud.}  
\begin{equation}
f_{NCRI}=1-\Theta/\Theta_{\rm rig} \; .
\label{eq:fs}
\end{equation}
In the case of a ring geometry, this quantity coincides  with the superfluid fraction defined in 
\cite{Leggett1970,Leggett1998}. In  Fig. \ref{fig:I-2d} we report our predictions for $f_{NCRI}$  as a 
function of the dimensionless parameter $\varepsilon_{dd}$.  This quantity exhibits a jump at the 
transition to the supersolid phase, which is much smaller than in the case of 
elongated trapping, followed by a further jump around $\varepsilon_{dd}=1.335$, corresponding to 
a change of the supersolid structure from the single-triangular cell (Fig. 3(c)) to the two 
triangular-cell (Fig. 3(d)) configuration. For larger values of $\varepsilon_{dd}$, the NCRI fraction  
 continues decreasing, the global behavior being  similar to the one of the superfluid  
fraction   calculated in periodic configurations as a function 
of $\varepsilon_{dd}$, both in the 1D \cite{Roccuzzo1} and in the 2D \cite{Pohl2019} case. In order 
to get a better insight on the rotational effects taking place in the supersolid phase we show in Fig.~\ref{fig:velocity} the velocity field 
${\bf v}({\bf r})=(\hbar/m)\nabla S({\bf r})$ of the rotating supersolid  ($\varepsilon_{dd}=1.347$). Despite the irrotational nature of the eGPE, the figure clearly reveals the rotational motion of 
the droplets through the superfluid, which reacts to the motion of the droplets. 

\begin{figure}
\includegraphics[width=\linewidth]{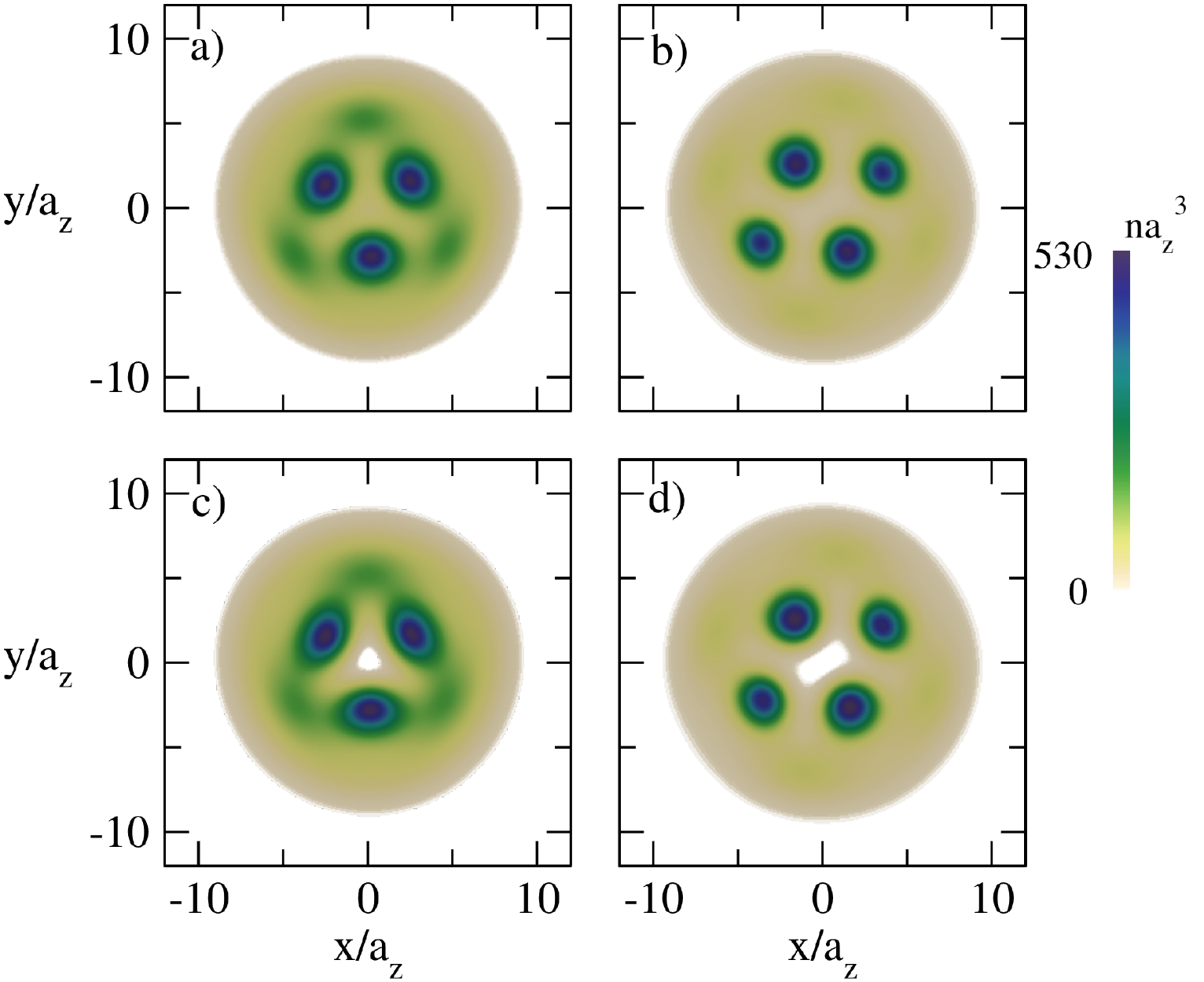}
\caption{Density plots of the ground state and vortical configuration: 
(a) and (c) in the single-triangular cell structure for $\varepsilon_{dd}=1.334$;
(b) and (d) in the case of a two-triangular cell structure obtained for $\varepsilon_{dd}=1.351$.}
\label{fig:vortex}
\end{figure}
The isotropic rotating configuration is very well suited to explore another important 
effect of superfluidity, i.e., the emergence of quantized vortices. Indeed we find that, at 
higher values of the angular velocity $\Omega$,  the supersolid is able to sustain a quantized 
vortex, thanks to the existence of an important superfluid component.  In Fig. \ref{fig:vortex} 
we show the 2D density profiles of a rotating supersolid configuration at frequency $\Omega=0.1\omega_x$ 
both in the single-triangular cell and the two-triangular cell structure case. The presence of the vortex is clearly revealed by the vanishing of the  density in the region of the vortex core 
(and - not shown - by the typical divergent behavior of the velocity field 
in the proximity of the center of the core), an effect directly 
measurable in future experiments. Remarkably, due to the small value of the density 
in the superfluid region and the vicinity of the surrounding droplets, the core of the vortex 
is large and deformed. The core deformation strongly depends on the structure of the droplets, 
being triangular-shaped and oblate in Fig. \ref{fig:vortex}(b) and (d), respectively. We find 
that the vortical solution becomes the ground state configuration for 
$\Omega > \Omega_c \sim 0.12 \omega_x$, i.e. for values of $\Omega$ significantly smaller than in 
the case of usual condensates \cite{BecBook2016}. The effect is the consequence of the small value 
of the density in the region where the vortex is formed. Furthermore we find that the jump in the 
angular momentum per particle, caused by the appearence of the vortex, is smaller than $\hbar$, 
reflecting the fact that the superfluid  fraction is smaller than $1$ in the supersolid phase. A 
more systematic discussion of the behavior of vortices in the supersolid phase will be 
the object of a future work.

In conclusion we have shown that supersolid  dipolar atomic gases confined in harmonic traps reveal important 
superfluid features. In the case of elongated configurations in the plane of rotation we have shown that the frequency of the scissors 
mode is a direct indicator of the effects of superfluidity on the moment of inertia, while in the case of isotropic trapping 
we have shown that, remarkably, the supersolid can host a quantized vortex whose core exhibits a characteristic deformed shape, 
caused by the presence of the surrounding droplets.  
Our theoretical predictions suggest that future measurements of the rotational effects will provide new light on the superfluid behavior of  these novel systems.

While completing this work we became aware of a very recent experimental work \cite{TanziScissor} reporting the measurement
of the scissors mode frequency in a dipolar supersolid and showing a very good agreement with our prediction.  

\paragraph*{Acknowledgments}
Stimulating discussions with Francesca Ferlaino, Giacomo Lamporesi,
Giovanni Modugno, Tilman Pfau and Luca Tanzi are acknowledged.
This  project  has  received  funding  from  the  European Union's Horizon 2020 research and innovation 
programme under grant agreement No.  641122 ``QUIC", from Provincia  Autonoma di Trento, the Q@TN initiative and
the  FIS$\hbar$ project  of  the  Istituto  Nazionale  diFisica Nucleare.

\bibliography{biblio-inertia.bib}

\end{document}